\documentclass[12pt,a4paper]{article}

\usepackage{color}
\usepackage{graphicx}
\usepackage{amsmath}
\usepackage{amsfonts}
\usepackage{amssymb}
\usepackage{subcaption} 
\usepackage{caption}
\usepackage[hidelinks]{hyperref}
\usepackage{mathrsfs}
\usepackage{enumerate}
\newcommand{\comment}[1]{}

\def\bsb{\boldsymbol}
\def\bea{\begin{eqnarray}}
\def\eea{\end{eqnarray}}
\def\be{\begin{equation}}
\def\ee{\end{equation}}

\newcommand{\mal}[1]{\mathcal #1}
\newcommand{\expect}[1]{\left\langle #1 \right\rangle}

\def\R{\mathcal R}
\def\O{\mathcal{O}}
\def\q{{\bsb q}}
\def\k{{\bsb k}}
\def\x{{\bsb x}}

\def\d{\partial}
\def\ep{\epsilon}

\def\mpl{M_{\rm pl}}

\def\zg{$\zeta\!-\!gauge$}
\def\rg{$\R\!-\!gauge$}

\begin{document}

\begin{center}
\Huge{\textbf{Solid Consistency}} \\
\end{center}
\vspace{0.1cm}

\begin{center}

\large{Lorenzo Bordin,$^{\rm a,b}$ Paolo Creminelli,$^{\rm c}$ Mehrdad Mirbabayi,$^{\rm c,d}$  Jorge Nore\~na$^{\rm e}$}
\\[0.5cm]

\small{
\textit{$^{\rm a}$ SISSA, via Bonomea 265, 34136, Trieste, Italy}}

\small{
\textit{$^{\rm b}$ INFN, National Institute for Nuclear Physics, Via Valerio 2, 34127 Trieste, Italy}}

\small{
\textit{$^{\rm c}$ Abdus Salam International Centre for Theoretical Physics\\ Strada Costiera 11, 34151, Trieste, Italy}}

\small{
\textit{$^{\rm d}$ Stanford Institute for Theoretical Physics, Stanford University, Stanford, CA 94305, USA}}

\small{
\textit{$^{\rm e}$ Instituto de F\'isica, Pontificia Universidad Cat\'olica de Valpara\'iso, Casilla 4059, Valpara\'iso, Chile}}

\vspace{.1cm}

\end{center}

\vspace{.8cm}

\hrule \vspace{0.1cm}
\noindent \small{\textbf{Abstract}\\ 
\noindent We argue that {\em isotropic} scalar fluctuations in solid inflation are adiabatic in the super-horizon limit. During the solid phase this adiabatic mode has peculiar features: constant energy-density slices and comoving slices do not coincide, and their curvatures, parameterized respectively by $\zeta$ and $\R$, both evolve in time.
 The existence of this adiabatic mode implies that Maldacena's squeezed limit consistency relation holds after angular average over the long mode. The correlation functions of a long-wavelength spherical scalar mode with several short scalar or tensor modes is fixed by the scaling behavior of the correlators of short modes, independently of the solid inflation action or dynamics of reheating.

\vspace{0.2cm}
\noindent
\hrule

\section{Introduction and Results}

Consistency relations (CRs) in single-field inflation are a consequence of adiabaticity: a long mode is locally unobservable and its effect can be removed by a coordinate redefinition \cite{Maldacena:2002vr,Creminelli:2004yq}. In the presence of additional fields, long-wavelength relative fluctuations (entropy modes) can be locally observed and CRs are violated. This common lore is challenged when one considers models of inflation with a different symmetry structure that cannot be described in the framework of the EFT of inflation \cite{Cheung:2007st}. In this paper we focus on the case of solid inflation \cite{Gruzinov,Endlich:2012pz} where the ``stuff" that drives inflation has the same symmetry as an ordinary solid. Here the situation is different from the usual case. In solids there is a single scalar excitation: the longitudinal phonon. However this mode is not adiabatic: the perturbation is anisotropic and this anisotropy is locally observable even at very long wavelengths. The absence of adiabaticity suggests at first sight that one cannot derive any CR.

This conclusion is too quick. The existence of CRs also with this different symmetry structure can be seen in this way. If one considers an isotropic perturbation of the solid, i.e.~a dilation or compression, this will be adiabatic since in solids there is a unique relation between the pressure and the energy density, $p(\rho)$. Since the solid experiences all states of compression as the universe expands, this perturbation cannot be locally distinguished from the unperturbed evolution. We are going to verify this statement in Section \ref{sec:linear} showing that an isotropic superposition of linear scalar modes is indeed adiabatic. This adiabatic mode is not standard: the two variables $\zeta$ and ${\cal R}$ do not coincide and they are both time dependent. This stems from the fact that the solids do not admit curved FRW solution, but only flat ones. 

The existence of adiabatic modes imply CRs for the variable $\zeta$. This does not happen for ${\cal R}$ since the diffeomorphism which removes the long mode cannot be written in terms of $\R$~in a model-independent way. In Section \ref{sec:consistency} we are going to verify the CRs in various cases, both when the short modes are inside the Hubble radius and outside. The conclusion is that, in models with the symmetry pattern of solid inflation, after reheating correlation functions satisfy the usual CRs once an average over the relative orientation between long and short modes has been done. We discuss the implications of this and open questions in Section \ref{sec:conclusions}.

Before proceeding, let us recall, following \cite{Endlich:2012pz}, that the dynamics of the solid is described in terms of three scalar fields $\phi^I$ which parametrise the position of the elements of the solid:
$\phi^I = x^I + \pi^I$\,.
The action can be written in terms of SO(3)-invariant objects built out of the matrix $B^{IJ} \equiv \partial_\mu\phi^I \partial^\mu \phi^J$. One can choose these invariants to be ($[\ldots]$ indicates a trace)
\be\label{XYZ}
X \equiv [B] \;,\qquad Y \equiv \frac{[B^2]}{[B]^2} \;, \qquad Z \equiv \frac{[B^3]}{[B]^3} \;.
\ee
So the action, at lowest order in derivatives and including gravity, is
\be
S = \int d^4x \sqrt{-g} \left[\frac{M_{Pl}^2}{2} R + F(X,Y,Z) \right].
\ee

\section{\label{sec:linear}Long Isotropic Perturbation}


\subsection{Proof of adiabaticity}

We want to show that, in solid inflation, a scalar perturbation becomes adiabatic once we average over the solid angle, i.e.~we take a superposition of scalar Fourier modes which is isotropic. We have to prove that this kind of perturbation, in the long-wavelength limit, can be brought back to the unperturbed solution via a suitable diffeomorphism.

We start from the so-called Spatially Flat Slicing Gauge (SFSG) which is defined as the gauge where the spatial part of the metric is only perturbed by tensor modes.  For the rest of this Section we will only be interested in scalar modes:  
\be
g_{ij} = a^2 \delta_{ij}\,, \hspace{2cm}  \phi^I = x^I + \pi^I\,.
\ee
The triplet $\pi^I$ consists of a scalar, $\pi_L$, plus a transverse vector, $\pi^I_T$, which we neglect in the following. The constraint equations give the lapse $N =1 + \delta N$ and the longitudinal part of the shift $N_L$ \cite{Endlich:2012pz}:\footnote{We use the notation: $\pi^i =\frac{\partial_i}{\sqrt{-\nabla^2}} \pi_L + \pi^i_T$ and analogously for $N^i$.}
\be
\label{SFSGconstraints}
\delta N = -\frac{a^2 \dot H}{k H} \frac{\dot\pi_L-\dot H \pi_L/H}{1-3 \dot H a^2/k^2} \;,\qquad N_L =  \frac{-3a^2 \dot H \dot\pi_L/k^2+ \dot H \pi_L/H}{1-3 \dot H a^2/k^2}\;.
\ee

In SFSG, the gauge-invariant variable $\zeta$ is given by
\be\label{zeta_of_pi_lineat}
\zeta = \frac{\d \pi}{3} + \O(\pi^2)\;,
\ee 
where $\partial\pi \equiv \partial_i \pi^i$.\footnote{Notice that when we expand around the background $\phi^I = x^I$ one has $\partial_i \phi^I = \delta_i^I$, so that there is no distinction between capital and lower-case spatial indeces.}
Therefore, by performing the following time diff,
\be \label{time_diff_zg}
x^0 \rightarrow x^0 + \xi^0(t,\x)\,, \ \ \  \mbox{with}\ \ \ \xi^0(t,\x) = \frac{1}{3H}\d \pi \,,
\ee
we go from SFSG to the \zg, defined by the condition $\delta \rho = 0$, where $\rho$ is the energy density. The spatial part of the metric now reads $g_{ij} = a(t)^2\ (1+2\, \zeta\!(t,\x))\ \delta_{ij}$, while one can write $\pi_L$ as a function of $\zeta$.
Using eq.s \eqref{SFSGconstraints} and \eqref{time_diff_zg}, one can verify that the expression of $\delta N$ in \zg\ is
\be
\label{deltaN}
\delta N = -\frac{k}{3}{\frac{d}{dt}\left( \frac{\pi_L}{H} \right)}\ \frac{1}{1-3 \dot H \frac{a^2}{k^2}}\;.
\ee
In the limit $k/aH \to 0$ the time-dependence of $\pi_L$ is slow-roll suppressed, see eq.~\eqref{wavefunctions}, therefore $\delta N \to 0$ on super-horizon scales. 

To reproduce the unperturbed FRW solution one has to eliminate the perturbation in the scalar fields $\phi^i$. This can be done by a redefinition of the spatial coordinates
\be\label{rescale}
x^i \rightarrow x^i + \xi^i(t,\x)\,, \ \ \  \mbox{with}\ \ \ \xi^i(t,\x) = - \pi^i(t,\x)\,.
\ee
Since now the scalars are unperturbed, it is natural to call this Unitary Gauge (UG).  In UG the shift vanishes on super-horizon scales
\be
\label{deltaNzeta}
N_{L} = -\frac{d}{dt}\left( \frac{\pi_L}{H} \right) \ \frac{H}{1-3\dot H \frac{a^2}{k^2}}\;.
\ee
In this gauge, for long wavelength, $\delta N = N_L = \pi^i =0$. However the spatial part of the metric is still perturbed
\be\label{spatial_metric_UG}
g_{ij} = a(t)^2\ (1+ 2 \zeta(t,\x) \delta_{ij} + \d_i\d_j \chi(t,\x))\,, \ \ \  \mbox{with} \ \ \  \chi(t, \x) = - {6}\ \d^{-2} \zeta(t, \x)\,.
\ee
The perturbation is purely anisotropic, i.e.~the volume is not perturbed because of the gauge condition $\delta\rho = 0$. Therefore if one considers a spherically symmetric superposition of scalar modes, the metric perturbations in eq.~\eqref{spatial_metric_UG} average to zero
\be
\int \frac{d^2 \hat \k}{4\pi} (2 \zeta_\k \delta_{ij} - k_i k_j \chi_\k) = \int \frac{d^2 \hat \k}{4\pi} (2 \zeta_\k \delta_{ij} - 6 \hat k_i \hat k_j \zeta_\k) = 0\;.
\ee
This shows that a spherically symmetric superposition of scalar modes is adiabatic. 

Notice that, if one works in \zg, the transformation that eliminates a long-wavelength mode and goes back to FRW is a rescaling of the spatial coordinates: this is quite similar to the standard case of single-field inflation. However here the rescaling is time-dependent and adiabaticity requires an average over directions. 
In the next Section we will see that the adiabaticity gives rise to CRs: the only difference with the standard case is that they hold only after the spherical average. (The time dependence of the rescaling is immaterial because one is usually interested in correlation functions at equal time.) 

Instead of using time-slices with $\delta\rho=0$, one could use slices that are orthogonal to the 4-velocity of the solid. The perturbation of the spatial part of the metric is called $\zeta$ in the first case and $\R$ in the second. Contrary to the usual case, in Solid Inflation the variables $\zeta$ and $\R$ differ even on super-horizon scales. At linear level
\be\label{R_of_zeta}
\R= \frac{1}{\ep H} \frac{\dot \zeta + \ep\, H\, \zeta}{1+k^2 /3 a^2 H^2\epsilon}\;.
\ee
Since the two slicings do not coincide, one needs a time-diff to go from one to the other. This is the difference of the two time diff.s to go from SFSG to \zg~and to \rg~respectively:
\be\label{time_shift_R_zeta}
\delta t _{\R\rightarrow\zeta} = \delta t_{\zeta} - \delta t _{\R} = \frac{\zeta}{H}-\frac{\R}{H} \simeq -\frac{\dot \zeta}{\epsilon H^2}\;,
\ee
where the last equation holds on super-horizon scales.
The property that $\delta N$ vanishes in \zg~on large scales, eq.~\eqref{deltaN}, will not hold in \rg. This means that to go from \rg~to the unperturbed FRW one has to supplement the rescaling of spatial coordinates with the time diff eq.~\eqref{time_shift_R_zeta}. As we will discuss in the next Section this implies that the CR for $\R$ will contain an extra piece: the time diff induces a piece involving the time-derivative of the short modes.

\subsection{Squeezed vs Super-Squeezed regimes}
In taking the long-wavelength limit $k \to 0$ one ends up in the regime $k \ll a H \epsilon^{1/2}$. However one expects that the adiabaticity arguments above hold whenever $k$ is comfortably outside the Hubble radius and in particular also in the intermediate regime $a H \gg  k \gg a H \epsilon^{1/2}$. This is indeed the case. For example in this regime the expression for the lapse in \zg~is
\be
|\delta N_\k| = \left|\frac{d}{dt}\left( \frac{\zeta_\k}{H} \right) \ \frac{1}{1-3 \dot H \frac{a^2}{k^2}}\right| < \left|\frac{d}{dt}\left( \frac{\zeta_\k}{H} \right) \ \frac{k^2}{a^2 H^2} \frac{1}{3 \dot H} \right|= \O\left( \frac{k^2}{a^2 H^2} \right)\,.
\ee
(Notice that in the inequality above we are {\em not} assuming that the term proportional to $\dot H$ in the denominator dominates.) The same argument works for the shift: also in the intermediate regime the physical difference with the unperturbed solution are suppressed when the mode is superhorizon.

Therefore we expect the CR to hold both in the intermediate and in the super-squeezed, $k \ll a H \epsilon^{1/2}$, regime. However, to get analytical results one is forced to expand the solution of the constraints in different ways in the two regimes and therefore one has to assume one of the two regimes. This also applies to the expression of the wavefunction, which can be expanded in the two limits to give \cite{Endlich:2012pz}

\be
\label{wavefunctions}
\pi_L(\tau, \k) \simeq \begin{cases}
\mal{B}_k \left( 1 + i c_L k \tau +\frac{1}{3} c_L^2 k^2 \tau^2 \right) e^{i c_L k \tau}\,,       & \text{if $|c_L k \tau| \ge \ep$\,,} \\
\mal B_k \left[1+\ep(1+c_L^2)\log(-c_L k \tau)\right] (-c_L k \tau_c)^{-5s/2 -\eta/2 -\ep} \,, & \text{if $|c_L k \tau| \le \ep$\,,}
\end{cases}
\ee
with   
\be
\mal B_k = -\frac{3}{2} \frac{H}{M_{Pl}\, c_L^{5/2}\, \ep^{1/2}} \frac{1}{k^{5/2}}\,.
\ee

\subsection{Adiabatic modes and the Weinberg theorem}

Solid Inflation is an interesting exception to many general theorems on cosmological perturbations. Weinberg \cite{Weinberg:2003sw,Weinberg:2008zzc} showed that, under quite general assumptions, one can always find an adiabatic mode which features \emph{identical} and \emph{time-independent}  $\zeta$ and $\R$ on super-horizon scales. In the Solid case, $\zeta$ and $\R$ are neither equal (see eq.~\eqref{R_of_zeta}) nor time-independent (both $\zeta$ and $\R$ have a slow-roll suppressed time-dependence on super-horizon scales). Of course by linearity these properties are not changed by the spherical average. In the original paper on Solid Inflation \cite{Endlich:2012pz} (see also \cite{Akhshik:2015rwa}) the authors addressed the issue of why a scalar Fourier mode does not comply with Weinberg analysis. The point is that the solid supports a large anisotropic stress, so that even in the long-wavelength limit the stress energy tensor remains anisotropic and thus locally distinguishable from the unperturbed solution: the mode is not adiabatic. The theorem \cite{Weinberg:2003sw,Weinberg:2008zzc} assumes the decay of the anisotropic stress for $k\rightarrow0$.

Here we are considering a spherical average of Fourier modes and the problem takes a somewhat different form. Indeed, as we discussed, the perturbation is now adiabatic, since the anisotropic stress averages to zero. However, this adiabatic mode is still different from the one of Weinberg. One can still check that the assumptions of the theorem do not hold: the $0i$ component of Einstein equations is not regular for $k\rightarrow0$, since $\delta u$ diverges in that limit. This does not allow to continue a homogeneous perturbation to a physical one at finite momentum.

This however looks rather technical. What is the physical reason why the adiabatic mode we are considering is different from the standard case? Why doesn't adiabaticity ensure that $\zeta$ is constant? In the standard case, the conservation of $\zeta$ and the relation $\zeta = \R$ can be understood from a linearized version of a spatially curved FRW. Neglecting short-wavelength perturbations and working at linear order, the curvature of a constant $\rho$ slice is given by
\be
^{(3)}R = - \frac{4}{a^2} \d^2 \zeta.
\ee
Since for a curved FRW the spatial curvature $\kappa = a^2\ {}^{(3)}R/6$ is constant, super-horizon $\zeta$ fluctuations better be time-independent. Moreover for a curved FRW the surfaces of constant density are perpendicular to the 4-velocity so we need $\zeta = \R$. The adiabatic mode we are discussing in Solid Inflation does not have these properties and this is related to the fact that one does not have curved FRW solution in this model.
\footnote{Curved FRW solutions are allowed if we change the internal metric in the Lagrangian, see \cite{Lin:2015cqa}. 
However, our argument for constancy of $\zeta$ in the standard case is based on the fact that curvature is a free parameter of the background solution. In the models discussed in \cite{Lin:2015cqa} curvature is uniquely fixed in terms of energy density, so we don't expect $\zeta$ to be conserved.}
This can be understood in terms of symmetries: the internal symmetries of the $\phi^I$ is isomorphic to the symmetries of flat Euclidean space. This allows to write flat FRW solutions, but not spatially curved solutions, which have a different group of isometries.\footnote{This also implies the usual curvature problem takes a somewhat different flavour in this class of models.}


\section{\label{sec:consistency}Angle-averaged Consistency Relations}

In \zg~a long mode averaged over the direction can be removed by a rescaling of the spatial coordinates. The derivation of the CR is very similar to the standard case, apart from the required angular average. In the case of the scalar 3-point function we get
\be
\label{mainzeta}
\int \!\frac{d^2\hat\q}{4\pi}\expect{\zeta_\q\zeta_\k\zeta_{-\k-\q}}'_{q \ll k} = -\frac{d \log k^3 P_\zeta(k)}{d\log k}  P_\zeta(q) P_\zeta(k) \,,
\ee
where here and in the following the prime indicates that a momentum-conserving delta function, $(2 \pi)^3 \delta(\sum_i {\bsb k_i})$, is dropped. We stress that this result, in the limit of exact scale-invariance when the RHS of eq.~\eqref{mainzeta} vanishes, was already discussed in \cite{Endlich:2013jia}.

\subsection{Check of the consistency relation for $\expect{\zeta\zeta\zeta}$}

Let us check the CR eq.~\eqref{mainzeta}. In Solid Inflation the quadratic action is $\O(\ep)$ while the cubic action is $\O(\ep^0)$. Thus the 3-point function is slow-roll {\em enhanced}, $f_{NL} \sim \expect{\zeta^3}/\expect{\zeta^2}^2 = \O(\ep^{-1})$ \cite{Endlich:2012pz}. Since the tilt of the 2-point function outside the horizon is $\O(\ep)$, a non-trivial (in the sense of non-zero) verification of eq.~\eqref{mainzeta} in this regime would require taking into account corrections to the leading bispectrum at second-order in slow-roll. This is quite challenging. We content ourselves with the first order correction: at this order the two sides of eq.~\eqref{mainzeta} should vanish when all the modes are outside the horizon. When the short modes are inside the horizon the scale-dependence of the spectrum is not slow-roll suppressed and the LHS of eq.~\eqref{mainzeta} should thus be non-zero.\footnote{For a discussion of CRs in standard inflation when the short modes are inside the horizon see \cite{Senatore:2012wy}.} The check will be done in the regime $k \gg a H \epsilon^{1/2}$ for all the modes.

To do this we compute the cubic Lagrangian in SFSG up to $\O(\ep)$, calculate the bispectrum and then transform to \zg. The $\O(\ep)$ corrections to the bispectrum of three super-horizon modes was studied in detail in \cite{Akhshik:2014bla}. Thus, we skip most of the technical steps. However, we identify a missing term in \cite{Akhshik:2014bla} that is important for the CR to work. The \emph{in-in} calculation up to this order consists of the sum of three pieces. Schematically,
\be\label{sketchy_scalar_3_point}
\expect{\zeta\zeta\zeta} \sim \mal L_{\O(1)}^{(3)} \times \pi(\tau, \k)_{\O(1)} + \mal L_{\O(1)}^{(3)} \times \pi(\tau,\k)_{\O(\ep)}+ \mal L_{\O(\ep)}^{(3)} \times \pi(\tau,\k)_{\O(1)} \,,
\ee
where $\mal L_{\O(\ep^n)}^{(3)}$ and $\pi(\tau,\k)_{\O(\ep^n)}$ are respectively the cubic Lagrangian and the wavefunctions evaluated at $n^{th}$ order in slow-roll. The leading cubic Lagrangian, $\mal L_{\O(1)}^{(3)}$, was calculated in \cite{Endlich:2012pz} (eq.~(D.2)). In Fourier space and in the squeezed limit it reduces to 
\be
\label{Lleading}
\mal L_{\O(1)}^{(3)}  \Big | _{\rm squeezed} = -\frac{8}{81} {F_Y} \left(1-3\cos^2(\theta)\right) \pi_{L,\q}\, \pi_{L,\k}\,\pi_{L,-\q-\k}\;,
\ee
where $\theta$ is the relative angle between the long and the short modes. The above expression gives zero when one takes the angular average. This means there is no contribution $\O(\ep^{-1})$ to the LHS of eq.~\eqref{mainzeta}. Notice that the cancellation after angular average holds independently of the explicit form of the wavefunctions. Therefore, the second term of eq.~\eqref{sketchy_scalar_3_point} vanishes and only the last term is relevant for checking the CR.

\paragraph{Cubic scalar Lagrangian at $\O(\ep)$.} At first look, expanding eq.~(D.1) of \cite{Endlich:2012pz} up to first order in slow-roll seems like a formidable task. Since $Y$ and $Z$ in \eqref{XYZ} are defined in such a way that they start from second order in perturbations, one has to expand
\be\label{expand}
\mal L_{\O(\ep)}^{(3)} =F_X \delta X^{(3)}+ F_{XX} \delta X^{(1)} \delta X^{(2)} + \frac{1}{6} F_{XXX} (\delta X^{(1)})^3
+ F_{XY} \delta X^{(1)} \delta Y^{(2)} + (F_Y \delta Y^{(3)})_{\O(\ep)} + (Y\leftrightarrow Z),
\ee
where subscripts on $F$ denote partial derivatives. However, there are several simplifications \cite{Akhshik:2014bla}. As we will see, one only needs the SFSG deformation matrix $B^{IJ}$ at zeroth order in slow-roll parameters. This allows neglecting $N^i$ and $\delta N$ (which are slow-roll suppressed in the regime we are considering):
\be\label{B}
B^{IJ} = \frac{1}{a^2}\left( \delta^{IJ} +\d^I\pi^J+ \d^J\pi^I +\d_k\pi^I\d_k\pi^J \right) -\dot\pi^I \dot\pi^J\, + \O(\ep)
\ee
Therefore, $\delta X$ terminates at quadratic order, apart from slow-roll suppressed corrections
\be
\delta X = \delta [B] = \frac{3}{a^2}(\frac{2}{3} \d\pi + \frac{1}{3}\d_i\pi_j \d_i\pi_j -\frac{a^2}{3} \dot\pi_i^2) +\O(\ep).
\ee
Given that derivatives of $F$ with respect to $X$ are slow-roll suppressed, there is no contribution from $F_X \delta X$ to $\O(\ep)$ cubic Lagrangian. Another simplification is that if at some order in perturbations the corrections to $[B^n]$ involve at most $m\leq n$ of the $B^{IJ}$ factors, then
\be\label{lema}
\delta \left(\frac{[B^n]}{[B]^n}\right) = \delta\left(\frac{[B^m]}{[B]^m}\right) ,\qquad \text{for all $n\geq m$ }.
\ee
This implies that 
\be
\delta Y^{(2)} = \delta Z^{(2)},
\ee
and since slow-roll corrections to $B^{IJ}$ start at $\O(\pi^2)$
\be
\delta Y_{\O(\ep)}^{(3)} = \delta Z_{\O(\ep)}^{(3)}.
\ee
Therefore, the following combinations appear in \eqref{expand}
\be
(F_Y+F_Z) \delta Y_{\O(\ep)}^{(3)},\qquad (F_{XY}+F_{XZ}) \delta X^{(1)} \delta Y^{(2)}.
\ee
However $F_Y + F_Z = \O(\ep)$ and $F_{XY}+F_{XZ} = \O(\ep^2)$, so these terms are negligible. Using
\be
F_{XX} = -\frac{a^4}{9}\ep F,\qquad F_{XXX} = \frac{2 a^6}{27} \ep F, \qquad F = -3 \mpl^2 H^2,
\ee
one gets
\bea\label{first_scalar_3_point}
\mal L_{\O(\ep)}^{(3)} &=& \ep M_{Pl}^2 H^2 a^3 \left[ \frac{2}{3} (\d\pi) \d_j\pi^k\d_j\pi^k - \frac{8}{27}(\d\pi)^3 \right] -\frac{2}{3} \ep M_{Pl}^2 H^2 a^2 (\d\pi)\dot\pi_i^2 \nonumber \\
&& +\frac{4}{27}(F_Y+F_Z)a^2 \left[ (\d\pi)\dot\pi_i^2 - 3 \d_i\pi^j\dot\pi^i\dot\pi^j \right]\,.
\eea
The terms with time derivatives in this equation are absent from eq.~(35) of \cite{Akhshik:2014bla}. Note that the appearance of the combination $F_Y+F_Z$ on the second line is a consequence of \eqref{lema} since time-derivatives appear in $B^{IJ}$ starting from quadratic order. The angular average of this term is zero, hence it does not contribute to CR while the last term on the first line does contribute.

\paragraph{Field redefinition.}
Since we are interested in the bispectrum of $\zeta$ at $f_{NL} = \O(1)$, we need to find the relation between $\zeta$ and $\pi$ at quadratic order and to zeroth order in $\ep$. We start from $B^{IJ}$ in SFSG given in \eqref{B}. The last term can be neglected because in the squeezed limit at least one of the two $\pi$'s will} be out of the horizon with a slow-roll suppressed time evolution. The assumption of spherical symmetry simplifies the expression,
\be
B^{IJ} = \delta^{IJ} X(t) = \frac{\delta^{IJ}}{a^2}\left( 1+\frac{2}{3} (\d\pi)+\frac{1}{9}{(\d\pi)}^2  \right)\,.
\ee
Now we perform the time diffeomorphism  that leads to the \zg~(the analogue of eq.~\eqref{time_diff_zg} but now at second order), where $X(t)$ takes its unperturbed value
\be
X(t+\xi^0(t,{\bsb x});\bsb x) = \bar X(t) = a^{-2}.
\ee
At non-linear order the spatial part of the \zg~ metric is defined as $g_{ij} = a^2 e^{2 \zeta}\delta_{ij}$. Therefore up to quadratic order in $\pi$, we obtain
\be\label{zeta_of_pi_quadratic}
\zeta = H \xi^0 = \frac{1}{3}\d\pi -\frac{1}{18}{(\d\pi)}^2+\frac{1}{9H}(\d\dot\pi)(\d\pi) + \O(({\d\pi})^3)\,.
\ee

One can now put together the \emph{in-in} computation, using the Lagrangian in the first line of eq.~\eqref{first_scalar_3_point}, with the definition of $\zeta$, eq.~\eqref{zeta_of_pi_quadratic}, to get
\be\label{scalar_3_point}
\int\! \frac{d^2\hat\q}{4\pi} \expect{\zeta_\q\zeta_\k\zeta_{-\q-\k}}'_{q \ll k} = \frac{1}{27} \int\! \frac{d^2\hat\q}{4\pi}\expect{(\d\pi)_\q(\d\pi)_\k(\d\pi)_{-\k-\q}}'_{q \ll k}-2P_\zeta(q)P_\zeta(k)+\frac{1}{H} P_\zeta(q) \dot P_\zeta(k)\,.
\ee
When all the modes are outside the horizon the last term on the RHS of eq.~\eqref{scalar_3_point} is slow-roll suppressed and can be neglected. A straightforward calculation shows that the other two terms cancel each other confirming that 
\be
\int \!\frac{d^2\hat\q}{4\pi}\expect{\zeta_\q\zeta_\k\zeta_{-\k-\q}}'_{q \ll k} = \O(\ep)\,,
\ee
as implied by eq.~\eqref{mainzeta}.\footnote{It is challenging to perform the calculation at next order in slow-roll, to test the CR. The cancellation after angular average in eq.~\eqref{scalar_3_point} can be seen only after the explicit  {\em in-in} integral (in contrast to what happens at leading order, eq.~\eqref{Lleading}). This means that, at higher order, we should compute time integrals involving the wavefunctions at $\O(\ep)$.} When the short modes are inside the horizon the cancellation between the first two terms on the RHS of eq.~\eqref{scalar_3_point} still holds, but now the last term is non-negligible since the time dependence is not slow-roll suppressed in this regime. One has
\be
\int \!\frac{d^2\hat\q}{4\pi}\expect{\zeta_\q\zeta_\k\zeta_{-\k-\q}}'_{q \ll k} = \frac{1}{H} P_\zeta(q) \dot P_\zeta(k) =-\frac{d \log k^3 P_\zeta(k,\tau)}{d\log k}  P_\zeta(q) P_\zeta(k,\tau) \,,
\ee   
where in the last passage we used that the power spectrum is of the form $P_\zeta = k^{-3} f(k \tau)$ as dictated by scale invariance, up to corrections of order slow-roll. The CR eq.~\eqref{mainzeta} is verified.

\subsection{Check of the consistency relation for $\expect{\zeta\gamma\gamma}$}
One novel feature of our analysis is that the (spherically averaged) adiabatic modes of solid inflation feature a time-dependent $\zeta$ outside the horizon. Since this time dependence arises at $\O(\ep)$ it would be nice to check the CR at this order. As discussed this is quite challenging for $\expect{\zeta\zeta\zeta}$, but it is doable for $\expect{\zeta\gamma\gamma}$, where the scalar mode is taken to be long. The leading term in $\expect{\zeta\gamma\gamma}$ is $\O(1)$ \cite{Endlich:2013jia}, so one just needs to do the calculation including the first-order slow-roll corrections. The leading Lagrangian has the form (see eq.~(A.7) of \cite{Endlich:2013jia})
\be
\mal L _{\O(\ep^0)}^{(3)} \propto -\frac13 (\partial \pi) \gamma_{ij} \gamma_{ij} +  \gamma_{ij} \gamma_{jk} \partial^k \pi^i.
\ee
It averages to zero in the squeezed limit $\pi_{L\,{\q\rightarrow0}}$ independently of the wavefunctions. Thus we do not need to consider the slow-roll corrections to the wavefunctions. We go directly to the computation of $\mal L _{\O(\ep)}^{(3)}$.

\paragraph{Cubic Lagrangian at $\O(\ep)$.}
There are two terms which contribute to $\mal L _{\O(\ep)}^{(3)}$. The first arises from the expansion of the function $F(X,Y,Z)$, while the second is from the Einstein-Hilbert action. The expansion of $F$ gives
\be \label{F_contr}
\mathcal L^{(3)}_{\O(\epsilon)}  \supset  F_X \delta X +F_Z \delta Z + \frac{F_{XX}}{2} {\delta X}^2 + F_{XY} \delta X \delta Y + F_{XZ} \delta X \delta Z 
\ee
where,
\bea
\delta X &=& a^{-2} \left[ 2 (\d\pi)- 2 \gamma_{ij}\d_i\pi^j +\frac{1}{2}{\gamma_{ij}}^2 +\gamma_{ij}\gamma_{jk}\d_k\pi^i \right] ,  \\
\delta Y &=&  \left[- \frac{4}{9} \gamma_{ij}\d_i\pi^j +\frac{1}{9}{\gamma_{ij}}^2 +\frac{2}{3}\gamma_{ij}\gamma_{jk}\d_k\pi^i  - \frac{2}{9}{\gamma_{ij}}^2(\d\pi) \right] ,  \\
\delta Z &=&  \left[- \frac{4}{9} \gamma_{ij}\d_i\pi^j +\frac{1}{9}{\gamma_{ij}}^2 +\frac{8}{9}\gamma_{ij}\gamma_{jk}\d_k\pi^i  - \frac{8}{27}{\gamma_{ij}}^2(\d\pi) \right] . 
\eea
One gets
\be
\mathcal L^{(3)}_{\O(\epsilon)} = -3 \epsilon a^2 M_{Pl}^2 H^2  \left[ \gamma_{ij}\gamma_{jk} \d_k\pi^j -\frac{1}{3} \gamma_{ij}^2 (\d\pi)\right] \;.
\ee
This gives zero after the angular average. The contribution which arises from the Einstein-Hilbert action (plus the appropriate boundary terms) is 
\bea\label{EH_term}
\mal L^{(3)}_{\O(\ep)} & \supset &  a^3 \left[ N \ {}^{(3)}R + N^{-1}(E_{ij}E^{ij}-E^2) \right]_{\O(\ep)\pi \gamma\gamma} \nonumber\\
&=& -\frac{a^2}{8} \delta N \left[ {\gamma'_{ij}}^2 +{(\d_l\gamma_{ij})}^2\right] -\frac{a^3}{4}\gamma_{ij}'\d_k\gamma_{ij}N^k\,.
\eea 
The {\em in-in} calculation can be done separately in the intermediate regime $a H \gg  q \gg a H \epsilon^{1/2}$ and in the super-squeezed regime $q \ll a H \epsilon^{1/2}$.
\footnote{Notice that eq.~\eqref{EH_term}, evaluated in the super-squeezed limit is $\O(1)$. But $\delta N,\ N_L \propto \dot\pi_L$ which has a slow-roll suppressed time dependence and so the final expression of the bispectrum is $\O(\ep)$. Notice also that, naively, the term proportional to the shift seems to cancel when one performs the angular average. However in the super-squeezed limit this term would give a bispectrum $\propto 1/q^4$ for $q \to 0$ (see the expression for $N^i$ in eq.~\eqref{SFSGconstraints}). This leading behaviour cancels, even before taking the angular average. The subleading contribution has the correct $1/q^3$ dependence and  does contribute to the angular average.} In both regimes the \emph{in-in} computation of the 3-point function gives 
\be\label{in-in_comp.}
\frac{1}{3}\int\! \frac{d^2\hat\q}{4\pi} \expect{({\d\pi})_\q\gamma_\k^s\gamma_{-\k}^s}_{q\ll k} = 2\ep\, P_\zeta(q)\,P_\gamma(k)\,.
\ee

\paragraph{Tensor modes at quadratic order in perturbations.}
The final expression of the bispectrum is given once one considers the contribution coming from the time diff that has to be performed to go from SFSG to \zg. This changes the tensor perturbations at quadratic level (see eq.~(A.8) of \cite{Maldacena:2002vr}). The interesting part (for us) is 
\be\label{tensor_time_diff}
\gamma_\zeta = \gamma_\pi + \frac{1}{H}\dot \gamma_\pi \frac{(\d\pi)}{3}\,,
\ee
where $\gamma_\zeta$  and $\gamma_\pi$ denote tensor perturbations respectively in \zg~and SFSG.
This adds a contribution to the bispectrum: $\expect{\zeta \gamma_\zeta\gamma_\zeta} = \expect{\zeta \gamma_\pi\gamma_\pi} +\frac{2}{3H}\expect{\zeta \d\pi \gamma_\pi \dot\gamma_\pi}$, which is given to leading order by
\be\label{tensor_redef}
\frac{1}{H}\, P_\zeta(q)\,\frac{d}{dt}P_\gamma(k) = -2\ep (1+c_L^2) P_\zeta(q)P_\gamma(k)\,.
\ee
Eqs.~\eqref{in-in_comp.} and \eqref{tensor_redef} give
\be\label{zggCR}
\begin{split}
\int\! \frac{d^2\hat\q}{4\pi} \expect{\zeta_\q\gamma_\k^s\gamma_{-\k}^s}_{q\ll k}= -2c_L^2 \ep P_\zeta(q)P_\gamma(k)
= -\frac{d \log k^3 P_\gamma(k)}{d\log k}  P_\zeta(q) P_\gamma(k)\,.
\end{split}
\ee
This is exactly what is predicted by the CR. In conclusion, this computation confirms that CRs after spherical average hold even at slow roll order, i.e. when the time dependence of the long mode cannot be neglected.

\subsection{Consistency relation for ${\cal R}$?}

There are two differences if one wants to use the variable $\R$ instead of $\zeta$.  First, as discussed above, starting from \rg~one needs an extra time diff to map a long mode into an unperturbed FRW. This changes the CR and introduces a time derivative of the short mode 2-point function. Moreover on super-horizon scale, in both squeezed and super-squeezed regimes, the relation between $\R$ and $\zeta$ depends on $c_L$ and therefore is non-universal,
\be
\R \simeq - c_L^2 \zeta.
\ee 
Hence, the spatial rescaling \eqref{rescale}, which is determined in terms of $\zeta$, will be model-dependent when written in terms of $\R$. This means that in this gauge we are not going to be able to write an explicit CR, i.e.~a model independent relation among correlation functions.

For instance, consider the squeezed limit of $\expect{\R\gamma\gamma}$.   One needs to go from \zg~to \rg~with the time diff $\delta_{\R\rightarrow\zeta}$, see eq.~\eqref{time_shift_R_zeta}. This implies
\bea
\int\! \frac{d^2\hat\q}{4\pi} \expect{\R_\q\gamma_\k^s\gamma_{-\k}^s}_{q\ll k}' &=& n_t \expect{\R_\q \zeta_{-\q}}' \expect{\gamma_\k\gamma_{-\k}}'+\frac{1}{\ep\,H^2}  \expect{\R_\q \dot\zeta_{-\q}}'  \frac{d}{dt} \expect{\gamma_\k\gamma_{-\k}}' = \nonumber \\
&=& 2\ep \left( 1-\frac{{(c_L^2+1)}^2}{c_L^2} \right) P_\R(q) P_\gamma(k)\,;
\eea
where we used the fact that, at first order in slow roll, $\dot \zeta_\q = - H (1+c_L^2)\, \ep\,   \zeta_\q$. This form of CR is not very useful since it contains parameters that make the relation model-dependent. However, after reheating $\zeta = \R$ and they are both time-independent. Therefore for observational purposes the CRs take the form of eqs.~\eqref{mainzeta} and \eqref{zggCR}.


\section{\label{sec:conclusions}Discussion and Outlook}
It is remarkable that in solid inflation one has evolution outside the horizon, even when the mode is spherically averaged and therefore adiabatic. The evolution is slow-roll suppressed during inflation, but it will become significant during reheating unless some extra assumption about this phase is made. (For instance the limit of instantaneous reheating was taken in \cite{Endlich:2012pz}.) Therefore one cannot even relate the normalization of the spectrum to the parameters during inflation, since the change during reheating may be significant. From this point of view, it is quite surprising that one can derive model-independent (and reheating independent) relations like eq.~\eqref{mainzeta}. 

Physically the angular-averaged consistency relations we found imply, within the symmetry pattern of solid inflation, that local (angle-independent) non-Gaussianity cannot be generated. (As in the usual case, the tiny non-Gaussianity in the squeezed limit, which is implied by the CR should be considered in some sense ``unobservable", see for example \cite{Pajer:2013ana}.) This however does not prevent large non-Gaussianity in the squeezed limit, as long as it vanishes after angular average. 

The consistency relation we discussed here is the analogue of the original Maldacena's CR. A natural question would be to look for other CRs, analogue of the conformal ones in standard inflation \cite{Creminelli:2012ed,Hinterbichler:2013dpa}. Another natural question is to try and apply these methods to other symmetry breaking patterns for instance Gauge-flation \cite{Maleknejad:2011jw} or supersolid inflation \cite{Bartolo:2015qvr,Ricciardone:2016lym}. In all these cases however there are multiple scalar excitations: similarly to standard multifield inflation, one does not expect any CR, unless further conditions are imposed. 

Finally, even though our explicit checks were limited to the lowest derivative solid action, the CRs are just based on symmetry considerations, and therefore they are robust when one considers higher derivative operators.

\subsection*{Acknowledgements}
We would like to thank Mohammad Akhshik, Alberto Nicolis, and Riccardo Penco for valuable discussions. 



\end{document}